\documentclass{elsart}
\usepackage{graphicx}

\def\ifm#1{\relax\ifmmode#1\else$#1$\fi}

\def\f{\ifm{\phi}}    

\def\ab{\ifm{\sim}}  
\def\x{\ifm{\times}}

\def\pt#1,#2,{\ifm{#1\x10^{#2}}}   
\def\epm{\ifm{e^+e^-}}
\renewcommand{\to}{\ensuremath{\rightarrow}}

\makeatletter
\newdimen\z@ \z@=0pt 
\newskip\z@skip \z@skip=0pt plus0pt minus0pt
\def\m@th{\mathsurround=\z@}
\def\ialign{\everycr{}\tabskip\z@skip\halign} 
\def\eqalign#1{\null\,\vcenter{\openup\jot\m@th              
  \ialign{\strut\hfil$\displaystyle{##}$&$\displaystyle{{}##}$\hfil
      \crcr#1\crcr}}\,}
\makeatother

\newcommand{\ep}{\mbox{$e^+$}}
\newcommand{\el}{\mbox{$e^-$}}
\newcommand{\ks}{\mbox{$K_S$}}
\newcommand{\kl}{\mbox{$K_L$}}
\newcommand{\kp}{\mbox{$K^+$}}
\newcommand{\km}{\mbox{$K^-$}}
\newcommand{\pip}{\mbox{$\pi^+$}}
\newcommand{\pim}{\mbox{$\pi^-$}}
\newcommand{\pio}{\mbox{$\pi^{0}$}}

\newcommand{\mpp}{\mbox{$M_{\pi\pi}$}}
\newcommand{\mpg}{\mbox{$M_{\pi\gamma}$}}

\newcommand{\dafne}{\mbox{DA$\Phi$NE}}
\newcommand{\br}{\mbox{BR}}

\newcommand{\wpo}{\mbox{$\omega\pio$}}
\newcommand{\wpc}{\mbox{$\pip\pim\pio\pio$}}
\newcommand{\wpn}{\mbox{$\pio\pio\gamma$}}
\newcommand{\eeto}{\mbox{$\ep\el\to\,\,$}}
\newcommand{\sqrts}{\mbox{$\sqrt{s}$}}

\newcommand{\aff}[2]{Dipartimento di Fisica dell'Universit\`a #1 e Sezione INFN, #2, Italy.}
\newcommand{\affd}[1]{Dipartimento di Fisica dell'Universit\`a e Sezione INFN, #1, Italy.}

\newcommand{\AmS}{{\protect\the\textfont2
  A\kern-.1667em\lower.5ex\hbox{M}\kern-.125emS}}

\begin{document}

\begin{frontmatter}

\title{\boldmath Study of the process \eeto\wpo\ with the KLOE detector}

%
\collab{The KLOE Collaboration}

\author[Na]{F.~Ambrosino},
\author[Frascati]{A.~Antonelli},
\author[Frascati]{M.~Antonelli},
\author[Frascati]{F.~Archilli},
\author[Roma3]{C.~Bacci},
\author[Karlsruhe]{P.~Beltrame},
\author[Frascati]{G.~Bencivenni},
\author[Frascati]{S.~Bertolucci},
\author[Roma1]{C.~Bini},
\author[Frascati]{C.~Bloise},
\author[Roma3]{S.~Bocchetta},
\author[Roma1]{V.~Bocci},
\author[Frascati]{F.~Bossi},
\author[Roma3]{P.~Branchini},
\author[Roma1]{R.~Caloi},
\author[Frascati]{P.~Campana},
\author[Frascati]{G.~Capon},
\author[Na]{T.~Capussela},
\author[Roma3]{F.~Ceradini},
\author[Frascati]{S.~Chi},
\author[Na]{G.~Chiefari},
\author[Frascati]{P.~Ciambrone},
\author[Frascati]{E.~De~Lucia},
\author[Roma1]{A.~De~Santis},
\author[Frascati]{P.~De~Simone},
\author[Roma1]{G.~De~Zorzi},
\author[Karlsruhe]{A.~Denig},
\author[Roma1]{A.~Di~Domenico},
\author[Na]{C.~Di~Donato},
\author[Pisa]{S.~Di~Falco},
\author[Roma3]{B.~Di~Micco},
\author[Na]{A.~Doria},
\author[Frascati]{M.~Dreucci},
\author[Frascati]{G.~Felici},
\author[Frascati]{A.~Ferrari},
\author[Frascati]{M.~L.~Ferrer},
\author[Frascati]{G.~Finocchiaro},
\author[Roma1]{S.~Fiore},
\author[Frascati]{C.~Forti},
\author[Roma1]{P.~Franzini},
\author[Frascati]{C.~Gatti},
\author[Roma1]{P.~Gauzzi},
\author[Frascati]{S.~Giovannella},
\author[Lecce]{E.~Gorini},
\author[Roma3]{E.~Graziani},
\author[Pisa]{M.~Incagli},
\author[Karlsruhe]{W.~Kluge},
\author[Moscow]{V.~Kulikov},
\author[Roma1]{F.~Lacava},
\author[Frascati]{G.~Lanfranchi},
\author[Frascati,StonyBrook]{J.~Lee-Franzini},
\author[Karlsruhe]{D.~Leone},
\author[Frascati]{M.~Martini},
\author[Na]{P.~Massarotti},
\author[Frascati]{W.~Mei},
\author[Na]{S.~Meola},
\author[Frascati]{S.~Miscetti},
\author[Frascati]{M.~Moulson},
\author[Frascati]{S.~M\"uller},
\author[Frascati]{F.~Murtas},
\author[Na]{M.~Napolitano},
\author[Roma3]{F.~Nguyen},
\author[Frascati]{M.~Palutan},
\author[Roma1]{E.~Pasqualucci},
\author[Roma3]{A.~Passeri},
\author[Frascati,Energ]{V.~Patera},
\author[Na]{F.~Perfetto},
\author[Lecce]{M.~Primavera},
\author[Frascati]{P.~Santangelo},
\author[Na]{G.~Saracino},
\author[Frascati]{B.~Sciascia},
\author[Frascati,Energ]{A.~Sciubba},
\author[Pisa]{F.~Scuri},
\author[Frascati]{I.~Sfiligoi},
\author[Frascati]{T.~Spadaro},
\author[Roma1]{M.~Testa},
\author[Roma3]{L.~Tortora},
\author[Roma1]{P.~Valente},
\author[Karlsruhe]{B.~Valeriani},
\author[Frascati]{G.~Venanzoni},
\author[Frascati]{R.Versaci},
\author[Frascati,Beijing]{G.~Xu}

\address[Frascati]{Laboratori Nazionali di Frascati dell'INFN, 
Frascati, Italy.}
\address[Karlsruhe]{Institut f\"ur Experimentelle Kernphysik, 
Universit\"at Karlsruhe, Germany.}
\address[Lecce]{\affd{Lecce}}
\address[Na]{Dipartimento di Scienze Fisiche dell'Universit\`a 
``Federico II'' e Sezione INFN,
Napoli, Italy}
\address[Pisa]{\affd{Pisa}}
\address[Energ]{Dipartimento di Energetica dell'Universit\`a 
``La Sapienza'', Roma, Italy.}
\address[Roma1]{\aff{``La Sapienza''}{Roma}}
\address[Roma3]{\aff{``Roma Tre''}{Roma}}
\address[StonyBrook]{Physics Department, State University of New 
York at Stony Brook, USA.}
\address[Beijing]{Permanent address: Institute of High Energy 
Physics of Academica Sinica,  Beijing, China.}
\address[Moscow]{Permanent address: Institute for Theoretical 
and Experimental Physics, Moscow, Russia.}

%
\begin{abstract}
%
Using $\sim$ 600 pb$^{-1}$ collected with the KLOE detector at \dafne, 
we have studied the production cross section of \wpc\ and \wpn\ final 
states in \ep\el\ collisions at center of mass energies between 1000 and 
1030 MeV.
By fitting the observed interference pattern around $M_\phi$ for both 
final states, we extract a measurement (preliminary) for 
the ratio $\Gamma(\omega\to\pio\gamma)/\Gamma(\omega\to\pip\pim\pio) = 
0.0934\pm 0.0022$.
Since these two final states represent the $98\%$ of the $\omega$ 
decay channels, we use unitarity to 
derive $\br(\omega\to\pip\pim\pio)= (89.94\pm 0.23)\%$ and
$\br(\omega\to\pio\gamma) = (8.40\pm 0.19)\%$.
Moreover, the parameters describing the \eeto\wpc\ reaction around $M_\phi$ 
are used to extract the branching fraction for the OZI and G-parity 
violating \f\to\wpo decay: $\br(\f\to\wpo) = (5.63\pm 0.70)\times 10^{-5}$. 
\end{abstract}
\begin{keyword}
$e^+e^-$ collisions \sep rare $\phi$ decays \sep VMD \sep OZI violation 
\sep Isospin violation
\end{keyword}

\end{frontmatter}

\section{Introduction}

In the energy region of few tens of MeV around $M_\phi$, the 
\eeto\pip\pim\pio\pio\ production cross section is largely dominated by 
the non-resonant processes $\eeto \rho/\rho' \to \omega \pio$. 
However, in a region closer to $M_\phi$, a contribution from the OZI and 
G-parity violating decay \f\to\wpo\ is expected.
This strongly suppressed decay can be observed only through the 
interference pattern with the previous reaction, which shows up as a dip 
in the production cross section as a function of the center of mass energy 
(\sqrts). A fit to the cross section energy dependence allows us to extract 
the \f\to\wpo\ branching fraction (BR).

There is a much more complicated interference scenario for the 
$\eeto\pio\pio\gamma$ channel. Here we expect contributions also from 
$\phi\to\rho\pi$ and $\phi\to S\gamma$ intermediate states, where
$S$ is a scalar meson.
In another paper \cite{Dalitz_piopiog} we have shown that at 
$\sqrts \sim M_{\phi}$ the interference between $\phi\to S\gamma$ and 
$\eeto\omega\pio$ events, evaluated by fitting the \mpp-\mpg\ Dalitz 
plot, is small.
Assuming this effect to be negligible to first order, a fit to the 
cross section interference pattern for this final state will nevertheless  
give information about the $\eeto\rho/\rho'\to\omega\pio$ 
process and the resonant decays $\phi\to\omega\pio$ and $\phi\to\rho\pio$. 
Comparing with the fit to the \pip\pim\pio\pio\ channel, the ratio 
$\Gamma(\omega\to\pio\gamma)/\Gamma(\omega\to\pip\pim\pio)$ can be
extracted.

\section{The KLOE detector}

The KLOE \cite{KLOE} experiment operates at \dafne\ \cite{DAFNE}, the 
Frascati $\phi$-factory. DA$\Phi$NE is an $e^+e^-$ collider running at 
a center of mass energy of $\sim 1020$~MeV, the mass of the 
$\phi$-meson. Equal-energy positron and electron beams collide at an 
angle of $\pi$-25 mrad, producing $\phi$-mesons nearly at rest.

The KLOE detector consists of a large cylindrical drift chamber, DC, 
surrounded by a lead-scintillating fiber electromagnetic calorimeter, EMC. 
A superconducting coil around the EMC provides a 0.52 T field. 
The drift chamber~\cite{DCH}, 4~m in diameter and 3.3~m long, has 12,582 
all-stereo tungsten sense wires and 37,746 aluminium field wires. The chamber 
shell is made of carbon fiber-epoxy composite and the gas used is a 90\% 
helium, 10\% isobutane mixture. These features maximize transparency to 
photons and reduce $\kl\to\ks$ regeneration and multiple scattering. The 
position resolutions are $\sigma_{xy}$\ab150 $\mu$m and $\sigma_z$\ab~2 mm. 
The momentum resolution is $\sigma(p_{\perp})/p_{\perp}\approx 0.4\%$. 
Vertices are reconstructed with a spatial resolution of \ab3~mm. 
The calorimeter~\cite{EMC} is
divided into a barrel and two endcaps, for a total of 88 modules, and covers 
98\% of the solid angle. The modules are read out at both ends by 
photo-multipliers, both in amplitude and time.  The readout granularity is 
\ab\,(4.4\x4.4)~cm$^2$, for a total of 2440 cells .  The energy deposits are 
obtained from the signals amplitude while the arrival times of particles and 
the positions in three dimensions are obtained from the time differences.
Cells close in time and space are grouped into a calorimeter 
cluster. The cluster energy $E$ is the sum of the cell energies. 
The cluster time $T$ and position $\vec{R}$ 
are energy weighed averages. Energy and time resolutions are $\sigma_E/E = 
5.7\%/\sqrt{E\ {\rm(GeV)}}$ and  $\sigma_t = 57\ {\rm ps}/\sqrt{E\ {\rm(GeV)}}
\oplus50\ {\rm ps}$, respectively.
The KLOE trigger \cite{TRG} uses both calorimeter and chamber information. 
In this analysis all the events are selected by the trigger calorimeter, 
requiring two energy deposits with $E>50$ MeV for the barrel and $E>150$ MeV 
for the endcaps.
A cosmic veto reject events where at least two outermost EMC layers are 
fired.
\section{\boldmath The \sqrts\ dependence of \eeto\wpc/\wpn\ cross sections}

As mentioned before, in the energy region below 1.4 GeV the 
$\pip\pim\pio\pio/\pio\pio\gamma$ production cross sections are dominated 
by the non-resonant process \\
$\eeto\rho/\rho'\to\wpo$. At $\sqrts \sim M_{\phi}$, 
the decay $\phi\to\wpo$ also contributes and interferes with 
the other processes. In the neutral channel there are also contributions
from $\phi\to S\gamma$ and $\phi\to\rho\pio$. 
The dependence of the cross section on the center of mass energy 
can be parametrized in the form \cite{phiwp2}:

\begin{equation} \label{eq:xsec}
\sigma(\sqrts) = \sigma_{0}(\sqrts)
\cdot\left|1-Z\frac{M_{\phi}\Gamma_{\phi}}{D_{\phi}}\right|
\end{equation}
where $\sigma_{0}(\sqrts)$ is the bare cross section for 
the non-resonant process, $Z$ is the complex interference parameter 
(i.e. the ratio between the $\phi$ decay amplitude and the non 
resonant processes), while $M_{\phi}$, $\Gamma_{\phi}$ and $D_{\phi}$
are the mass, the width and the inverse propagator of the $\phi$ meson 
respectively.
The non-resonant cross section in this energy range increases linearly
with \sqrts. A model independent parametrization \ref{eq:xsec_bare}will 
be used in this paper.
\begin{equation} \label{eq:xsec_bare}
\sigma_{0}(\sqrts) = \sigma_{0} + \sigma' (\sqrts - M_\phi)
\end{equation}
%

\section{Data analysis}

All the available statistics collected at the $\phi$ peak in 2001--2002 
data-taking periods, corresponding to 450 pb$^{-1}$, has been analyzed.
Moreover four scan points (1010 MeV, 1018 MeV, 1023 MeV and 1030 MeV) of 
$\sim 10\ {\rm pb}^{-1}$ each and the off-peak data (1000 MeV) acquired
in 2005-2006 have been included in this analysis. All runs are grouped in 
center of mass energy bins of 100 keV.

\subsection{$\eeto\omega\pio\to\wpc$}
In the \pip\pim\pio\pio\ analysis,
data are filtered by selecting events with the expected final state 
signature: two tracks connected to a vertex inside a small cylindrical 
fiducial volume around the Interaction Point (IP) and four neutral clusters 
in the prompt Time Window (TW), defined as
$|T_{\gamma}-R_{\gamma}/c|<MIN(3.5\cdot\sigma_T,2\mbox{ ns})$.
To minimize contamination from machine background, we require also a minimal 
energy for the clusters (10 MeV) and a minimal angle with respect to the 
beam line (\ab 23$^{\circ}$). 
\begin{table}[!hb]
\caption{Background channels for \wpo\to\wpc. Signal over background 
ratios after acceptance selection and $\chi^2_{\rm Kfit}$ cut are 
reported for events collected at $\sqrts \sim M_{\phi}$.}
\label{tab:wpcbkg}
\newcommand{\m}{\hphantom{$0$}}
\renewcommand{\tabcolsep}{0.6pc} 
\renewcommand{\arraystretch}{1.2} 
\begin{tabular}{@{}lcc} \hline\hline
Channel          &  S/B (acc)    &  S/B ($\chi^2_{\rm Kfit}$ cut)\\ \hline
$\ks\kl$         &  1            &   10               \\
$\kp\km$         &  10           &   60               \\
$\rho\pi$        &  30           &  200               \\
$\eta\gamma$     &  20           &  800  \\ \hline\hline
\end{tabular}
\end{table}
A kinematic fit requiring total four-momentum conservation and time of 
flight (TOF) for photons improves the energy resolution. 
The resulting $\chi^2$ ($\chi^2_{\rm Kfit}$) is used to select a 
signal enriched ($\chi^2_{\rm Kfit}<50$), S$_{\rm evts}$ , and a 
background dominated ($\chi^2_{\rm Kfit}>50$), B$_{\rm evts}$, samples.
The signal analysis efficiency in the S$_{\rm evts}$ sample has
been evaluted by Montecarlo (MC). The resulting value 
$\varepsilon \sim 40\%$ is dominated by the acceptance requirements and 
has a small dependence as a function of \sqrts.

The background channels are listed in Tab.\ \ref{tab:wpcbkg}. The main
contribution comes from $\phi\to\ks\kl\to\pip\pim\pio\pio$ and 
$\phi\to K^+K^-$ with $K^\pm\to\pi^\pm\pio$, which have the same final 
state. The first one has also a comparable production cross section 
with respect to the signal at the $\phi$ peak.
The other two background components ($\phi\to\eta\gamma$ with 
$\eta\to\pip\pim\pio$, and $\phi\to\pip\pim\pio$) mimic the final state 
signature because of additional clusters due to accidental coincidence of
machine background events and/or shower fragments (splitting).
In the signal enriched region, the expected contamination at 
$\sqrts\sim M_\phi$ is $\sim 12\%$.
\begin{figure}[ht]
\begin{center}
\includegraphics[width=1.\textwidth]{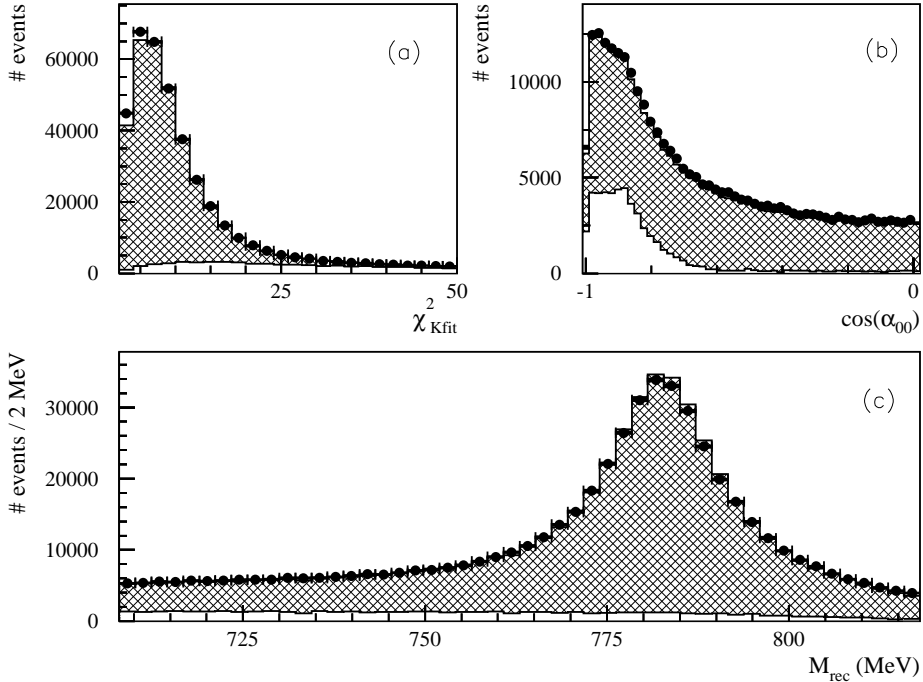}
\end{center}
\caption{Data-MC comparison for \pip\pim\pio\pio\ signal enriched 
distribution using events taken at 1019.75 MeV : (a) $\chi^2_{\rm Kfit}$
(Ndf=8); (b) cosine of the angle between reconstructed \pio's; 
(c) \pio\ recoil mass. 
Black dots are data, while hatched and white histograms are MC signal 
and background shapes respectively, weighted by our fit results.}
\label{fig:count}
\end{figure}

The signal counting on data is performed for each \sqrts\ bin by fitting the 
reconstructed \pio\ recoil mass ($M_{rec}$) distribution for both 
S$_{\rm evts}$ and B$_{\rm evts}$ samples with MC signal and background shapes.
The fit procedure is based on a likelihood function which takes into 
account both data and MC statistics.
In Fig.~\ref{fig:count}, data-MC comparison of few relevant distributions 
for the most populated energy bin is shown.

The results are summarized in Tab.~\ref{tab:count} where the signal counts,
the $\chi^2$ of the fit and the visible cross section are reported for all
\sqrts\ bins. A preliminary estimate of the systematic error, dominated by
tracking and vertexing efficiency, has been added to the 
$\sigma^{4\pi}_{\rm vis}$ error.

\begin{table}[!htb]
\caption{Signal counting, $\chi^2$ of the fit and visible cross section 
for $\ep\el\to\wpc$ events.
The errors on $\sigma^{4\pi}_{\rm vis}$ contains a relative systematic 
error contribution of 1.8\%.}
\label{tab:count}
\newcommand{\m}{\hphantom{$0$}}
\renewcommand{\tabcolsep}{0.6pc} 
\renewcommand{\arraystretch}{1.2} 
\begin{tabular}{@{}lccc} \hline\hline
\sqrts\ (MeV) & N$^{4\pi}\pm\delta_N$     & $\chi^2/{ndf}$ & $\sigma^{4\pi}_{\rm vis} \pm \delta_{\sigma}$  \\ 
\hline
1000.10      & $199099 \pm 1276$  &    1.09           &  $5.75\pm 0.11$  \\
1009.90      & $ 26379 \pm  255$  &    1.08           &  $6.46\pm 0.19$  \\
1017.20      & $ 16720 \pm  184$  &    1.06           &  $5.67\pm 0.12$  \\
1018.14      & $ 22824 \pm  199$  &    0.95           &  $5.76\pm 0.12$  \\
1019.19      & $  7851 \pm  112$  &    0.88           &  $5.40\pm 0.13$  \\
1019.45      & $ 59738 \pm  460$  &    1.10           &  $5.64\pm 0.11$  \\
1019.55      & $ 96610 \pm  641$  &    1.05           &  $5.81\pm 0.11$  \\
1019.65      & $175734 \pm 1261$  &    1.10           &  $5.74\pm 0.11$  \\
1019.75      & $336385 \pm 2271$  &    1.04           &  $5.86\pm 0.11$  \\
1019.85      & $264184 \pm 2061$  &    1.06           &  $6.05\pm 0.12$  \\
1019.95      & $ 36999 \pm  611$  &    1.00           &  $6.16\pm 0.15$  \\
1020.05      & $ 18358 \pm  433$  &    1.01           &  $6.10\pm 0.18$  \\
1020.15      & $  7293 \pm  356$  &    0.99           &  $6.06\pm 0.32$  \\
1020.41      & $  9067 \pm  222$  &    0.92           &  $6.08\pm 0.19$  \\
1022.09      & $ 19307 \pm  242$  &    1.06           &  $7.08\pm 0.16$  \\
1022.98      & $ 29995 \pm  265$  &    0.93           &  $7.52\pm 0.16$  \\
1029.97      & $ 35125 \pm  489$  &    1.01           &  $8.03\pm 0.22$  \\ 
\hline \hline
\end{tabular}
\end{table}

\subsection{$\eeto\omega\pio\to\wpn$}

The acceptance selection for $\pio\pio\gamma$ events requires five 
neutral clusters with $E_\gamma \ge 7$ MeV and a polar angle 
$|\cos\theta_\gamma| < 0.92$ in the prompt Time Window.
After applying a first kinematic fit (Fit1) imposing total 4-momentum 
conservation, photons are paired to \pio's, by minimizing a $\chi^2$ 
built using the invariant mass of the two $\gamma\gamma$ pairs.
A second kinematic fit (Fit2) imposes also constraints on the \pio\ 
masses.

\begin{figure}[!hb]
\begin{center}
\includegraphics[width=.9\textwidth]{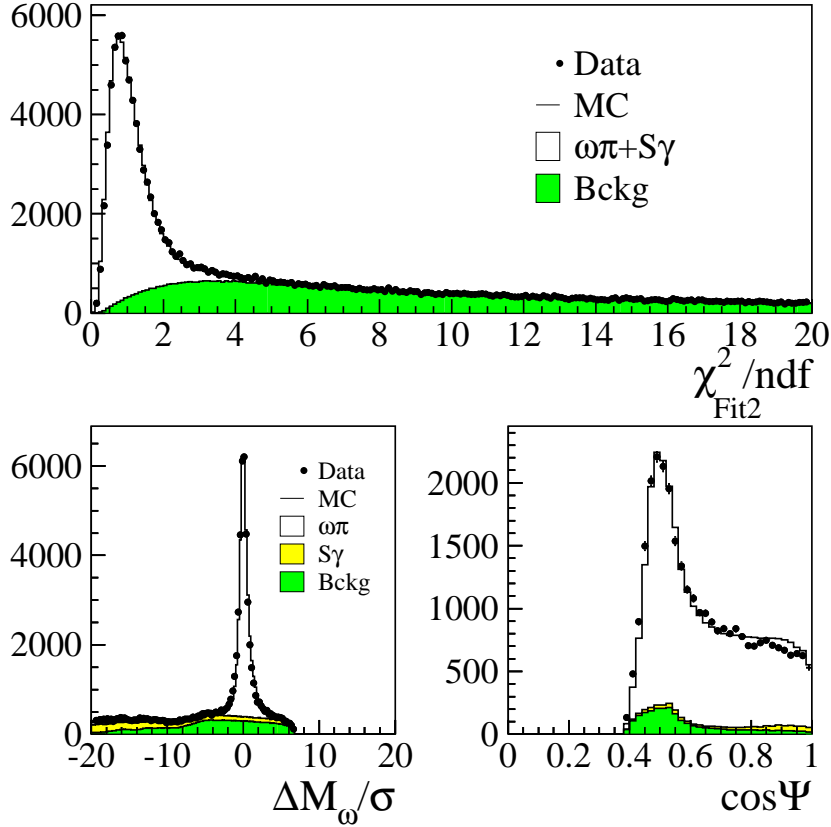}
\end{center}
\caption{Data-MC comparison for $\pio\pio\gamma$ events taken at 1019.75 MeV. 
Top: normalized $\chi^2$ of the second kinematic fit after acceptance cuts.
Bottom: $\pio\gamma$ invariant mass (left), and $\cos\psi$ distribution 
after cutting on $M_{\pi\gamma}$ (right).
In the upper panel all the background is grouped together, while in the 
lower ones the $\phi\to S\gamma$ contribution is shown alone.}
\label{Fig:DataMC_wpn}
\end{figure}

The background with final state different from $\pio\pio\gamma$ is 
rejected by requiring $\chi^2_{\rm Fit2}/{\rm Ndf}\le 5$ and
$\Delta M_{\gamma\gamma}= 
|M_{\gamma\gamma}-M_{\pi}| \le 5\,\sigma_{\gamma\gamma}$, where 
$M_{\gamma\gamma}$ and $\sigma_{\gamma\gamma}$ are evaluated using 
the photon momenta from Fit1. 
After these cuts the remaining sample is dominated by 
$\epm\to\omega\pio\to\pio\pio\gamma$ and $\phi\to S\to\pio\pio\gamma$
events. Signal is then selected neglecting the interference 
between the two processes and cutting on the intermediate state mass.
Since $M_{\pi\gamma}$ the closest mass to $M_{\omega}$ of the two 
$\pio\gamma$ combinations, only events satisfying
$|M_{\pi\gamma}-M_{\omega}|<3\,\sigma_{M_\omega}$ are retained.
In Tab.\ \ref{Tab:Bckg_wpn}, the background channels are listed together 
with the S/B ratio before and after the application of the whole analysis 
chain. The residual 10\% background contamination comes predominantly 
from $\phi\to\eta\gamma\to\pio\pio\pio\gamma$ events where two photons
are lost or merged.

In Fig.\ \ref{Fig:DataMC_wpn} data-MC comparison for events in the most
populated \sqrts\ bin is shown. The $\psi$ variable is the minimum angle 
between the photon and the \pio 's in the di-pion rest frame.
A good agreement is observed both after acceptance selection and after
applying analysis cuts.

The overall analysis efficiency for the identification of the signal is 
evaluated by applying the whole analysis chain to signal MC events: 
$\varepsilon_{\pi\pi\gamma} \sim 40\%$, almost flat in \sqrts.
The value obtained for each bin, together with the corresponding integrated 
luminosity, has been applied to the signal counting to obtain the
visible cross section. Results are summarized in Tab.\ \ref{Tab:xsec_wpn};
errors include statistics and background subtraction only.

\begin{table}[!ht]
  \caption{Background channels for $\epm\to\omega\pio\to\pio\pio\gamma$. 
  The signal over background ratio before and after the application of 
  the analysis cuts is reported for events collected at $\sqrts\sim M\phi$.}
\label{Tab:Bckg_wpn}
\newcommand{\m}{\hphantom{$0$}}
\renewcommand{\tabcolsep}{0.6pc} 
\renewcommand{\arraystretch}{1.2} 
\begin{tabular}{@{}lcc} \hline\hline
Background        & S/B (no cuts)   &   S/B (all cuts)\\ \hline
$\phi\to S\gamma\to\pio\pio\gamma$               & 1.5  &   35 \\
$\phi\to\eta\pio\gamma\to\gamma\gamma\pio\gamma$ & 5.4  &  120 \\
$\phi\to\eta\gamma\to\pio\pio\pio\gamma$         & 0.04 &   15 \\
$\phi\to\eta\gamma\to\gamma\gamma\gamma$         & 0.04 &  380 \\
$\phi\to\pio\gamma$                              & 0.13 & 2840 \\
\hline\hline
\end{tabular}
\end{table}

\begin{table}
  \caption{Signal counting and visible cross section for $\ep\el\to\wpn$ 
events.}
\label{Tab:xsec_wpn}
\newcommand{\m}{\hphantom{$0$}}
\renewcommand{\tabcolsep}{0.6pc} 
\renewcommand{\arraystretch}{1.2} 
\begin{tabular}{@{}lcc} \hline\hline
\sqrts\ (MeV) & N$^{\pi\pi\gamma}\pm\delta_N$  & $\sigma^{\pi\pi\gamma}_{\rm vis} \pm \delta_{\sigma}$  \\ 
\hline
 1000.10  &  $ 5523\pm  75$  &  $0.540\pm 0.007$  \\
 1009.90  &  $ 2445\pm  50$  &  $0.607\pm 0.012$  \\
 1017.15  &  $  831\pm  30$  &  $0.593\pm 0.020$  \\
 1017.25  &  $  680\pm  28$  &  $0.578\pm 0.022$  \\
 1018.20  &  $ 2088\pm  50$  &  $0.554\pm 0.013$  \\
 1019.35  &  $  273\pm  18$  &  $0.547\pm 0.036$  \\
 1019.45  &  $ 4911\pm  79$  &  $0.514\pm 0.008$  \\
 1019.55  &  $ 7693\pm 100$  &  $0.510\pm 0.006$  \\
 1019.65  &  $14788\pm 141$  &  $0.532\pm 0.005$  \\
 1019.75  &  $27556\pm 199$  &  $0.530\pm 0.004$  \\
 1019.85  &  $20927\pm 170$  &  $0.529\pm 0.004$  \\
 1019.95  &  $ 2869\pm  60$  &  $0.528\pm 0.011$  \\
 1020.05  &  $ 1475\pm  43$  &  $0.536\pm 0.015$  \\
 1020.15  &  $  577\pm  26$  &  $0.536\pm 0.024$  \\
 1020.45  &  $  542\pm  26$  &  $0.524\pm 0.024$  \\
 1022.25  &  $  996\pm  33$  &  $0.639\pm 0.021$  \\
 1022.35  &  $  415\pm  21$  &  $0.661\pm 0.034$  \\
 1022.95  &  $ 2574\pm  53$  &  $0.646\pm 0.013$  \\
 1029.95  &  $ 3233\pm  57$  &  $0.751\pm 0.013$  \\ \hline\hline
\end{tabular}
\end{table}

\section{Fit results and $\omega$ branching ratios extraction}

\begin{figure}[!ht]
\begin{center}
\includegraphics[width=1.\textwidth]{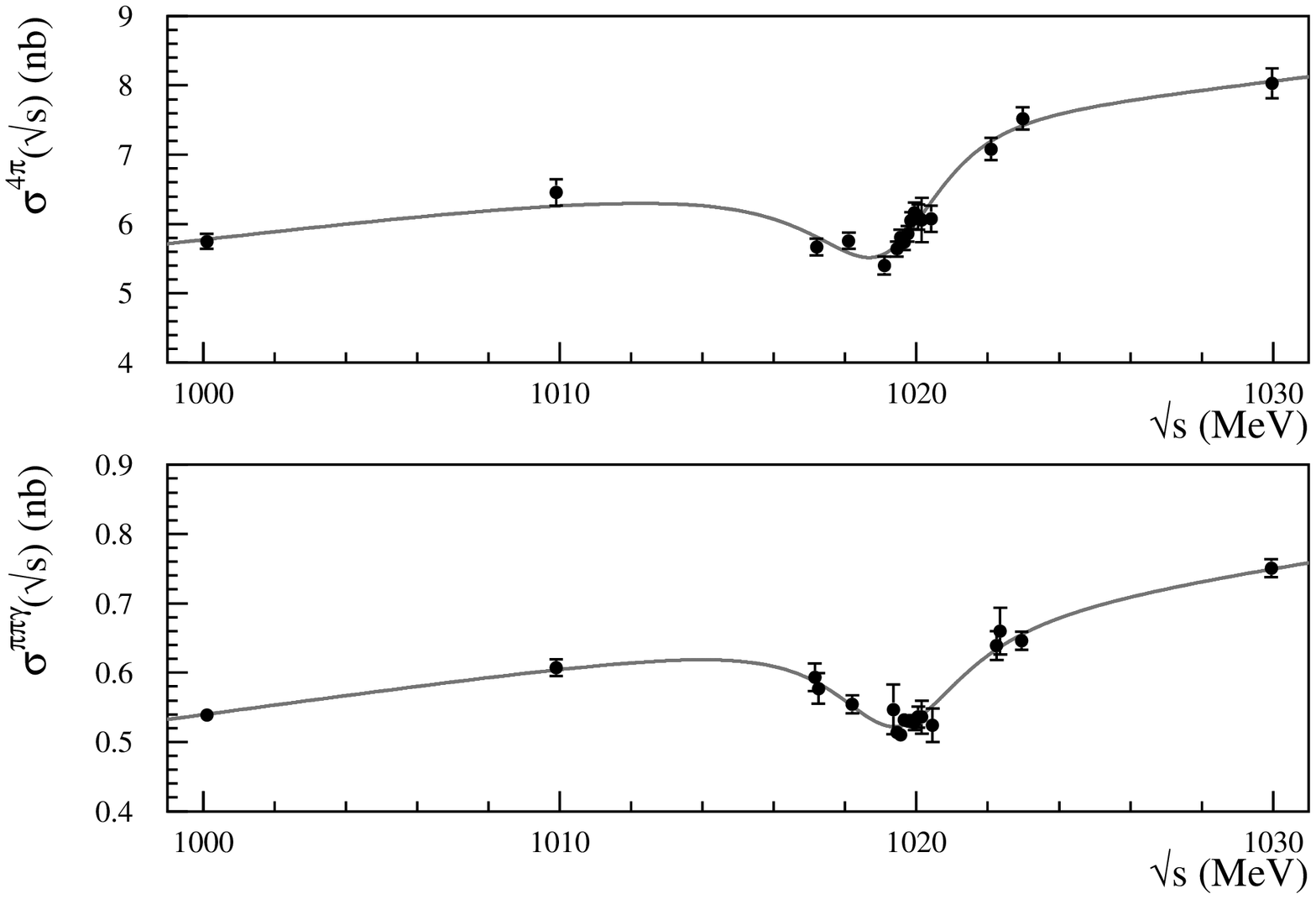}
\end{center}
\caption{Cross section fit results for the \eeto\wpc\ (top) and
\eeto\wpn\ (bottom) channels. 
Black dots are data, solid line is the resulting fit function.}
\label{fig:cross}
\end{figure}

The measured values of visible cross section, shown in 
Tab.~\ref{tab:count} and \ref{Tab:xsec_wpn}, are fitted with 
the parametrization (\ref{eq:xsec}), convoluted with a radiator function \cite{RAD}. 
The free parameters are: 
$\sigma_{0}^{i}$, $\Re(Z_{i})$, $\Im(Z_{i})$ and $\sigma'_{i}$,
where $i$ is the $4\pi$ or $\pi\pi\gamma$ final state.
In Fig.~\ref{fig:cross} data points with the superimposed fit function are 
shown for both channels. The preliminary values for the extracted parameters 
are reported in Tabs.~\ref{tab:fitres}. The resulting 
$\chi^2/Ndf$ are 12.8/15 ($P(\chi^2)=62\%$) for the fully neutral channel and 
13.4/13 ($P(\chi^2)=42\%$) for the other one.

\begin{table}[!htb]
\caption{Fit results for the $\eeto\wpc$ cross section (left)
and for $\eeto\wpn$ cross section (right).}
\label{tab:fitres}
\newcommand{\m}{\hphantom{$-$}}
\renewcommand{\tabcolsep}{0.6pc} 
\renewcommand{\arraystretch}{1.2} 
\begin{center}
\begin{tabular}{@{}lr}
\begin{tabular}{@{}ll} \hline\hline
\multicolumn{2}{l}{Parameter ($\eeto\wpc$)}              \\ \hline
$\sigma_0^{4\pi}$ (nb)      &   \m 8.12 $\pm$ 0.14       \\
$\Re(Z_{4\pi})$             &  \m$ 0.097 \pm 0.012$      \\
$\Im(Z_{4\pi})$             &  $-0.133 \pm 0.009$        \\
$\sigma'_{4\pi}$ (nb/MeV)   &\m  $ 0.072 \pm 0.008$      \\\hline \hline
\end{tabular}
~~~
&
~~~
\begin{tabular}{@{}ll} \hline\hline
\multicolumn{2}{l}{Parameter ($\eeto\wpn$)}                  \\ \hline
$\sigma_0^{\pi\pi\gamma}$ (nb)      &   \m 0.776 $\pm$ 0.012      \\
$\Re(Z_{\pi\pi\gamma})$             &  \m$ 0.013 \pm 0.013$  \\
$\Im(Z_{\pi\pi\gamma})$             &  $ -0.155 \pm 0.007$   \\
$\sigma'_{\pi\pi\gamma}$ (nb/MeV)   &  $ 0.0079 \pm 0.0006$  \\\hline \hline
\end{tabular}
\\
\end{tabular}
\end{center}
\end{table}

From the two previous measurements we obtain:
\begin{equation}
  \frac{\sigma_0(\omega\to\pio\gamma)}{\sigma_0(\omega\to\pip\pim\pio)} = 
  0.0956\pm 0.0022
\end{equation}
Taking into account the phase space difference between the two decays 
\cite{phiwp2}, the ratio of the partial widths can be extracted:
\begin{equation}
  \frac{\Gamma(\omega\to\pio\gamma)}{\Gamma(\omega\to\pip\pim\pio)} = 
  0.0934\pm 0.0021
\end{equation}

\begin{figure}[!ht]
\begin{center}
\includegraphics[width=0.7\textwidth]{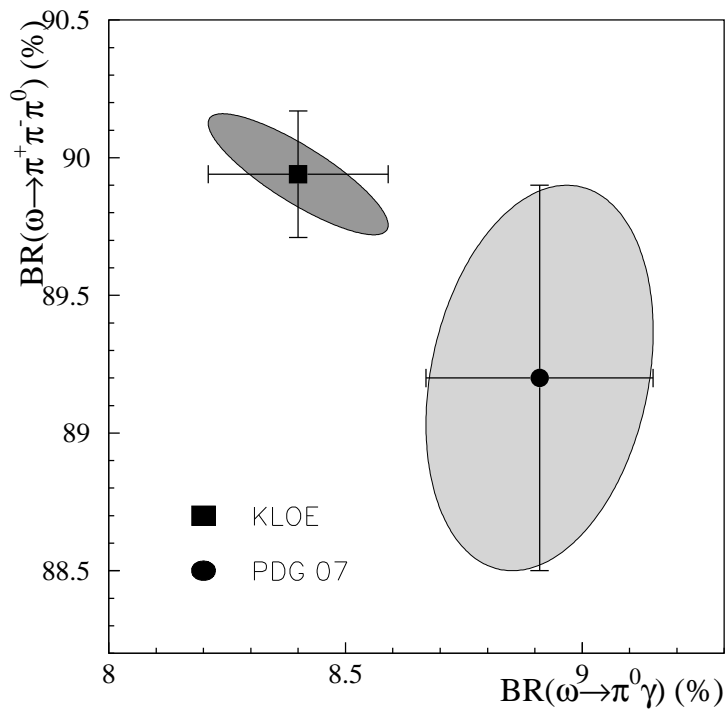}
\end{center}
\caption{Branching fraction for the two main $\omega$ decay channels. 
The black square is the KLOE fit result, while the black dot is the PDG
constrained fit result. The shaded regions are the 68\% C.L. }
\label{fig:wbrfit}
\end{figure}

Since these two final states the $98\%$ of the $\omega$ decay channels, we use 
the $\Gamma(\omega\to\pio\gamma)/\Gamma(\omega\to\pip\pim\pio)$ ratio and the 
sum of averages of the existing BR measurements on rarest decays \cite{PDG07} 
to extract the main $\omega$ branching fractions, imposing the unitarity relation:

\begin{eqnarray}
  BR(\omega\to\pip\pim\pio) & = & (89.94\pm 0.23)\% \\
  BR(\omega\to\pio\gamma)   & = & ( 8.40\pm 0.19)\%
\end{eqnarray}

with a correlation of 82\%. Comparison between our evaluation and the 
values in PDG \cite{PDG07} is shown in Fig.\ \ref{fig:wbrfit}. 

\section{BR($\phi\to\omega\pio$) evaluation}

The measured $\sigma_0^{4\pi}$ and $Z_{4\pi}$ paramters of the \wpc\ final 
state are related to the BR(\f\to\wpo) through the relation:
\begin{equation}
BR(\phi\to\wpo) = \frac{\sigma_0(m_{\phi})|Z_{4\pi}|^2}{\sigma_{\phi}},
\end{equation}
where $\sigma_0(m_{\phi})$ is the total cross section of the $\eeto\omega\pio$ 
process and $\sigma_{\phi}$ is the peak value of the production cross section 
for the \f\ resonance. 

Using the parameters obtained from the \wpc\ analysis, the $\Gamma_{ee}$ 
measurement from KLOE \cite{GeeKLOE} for the evaluation of $\sigma_{\phi}$,
and our value for  $BR(\omega\to\pip\pim\pio)$ we extract:

\begin{equation}
  BR(\phi\to\wpo) = (5.63\pm 0.70)\times 10^{-5}
\end{equation}
in agreement with the previous measurement from the SND experiment 
\cite{phiwp2}.

%

\end{document}